\documentclass[11pt,twoside]{article}

\usepackage{asp2006}
\usepackage{epsf}

\begin{document}
\title{Star Formation Properties and Dynamics of Luminous Infrared Galaxies 
with Adaptive Optics}   
\author{Petri V\"ais\"anen$^1$, Seppo Mattila$^2$, Stuart Ryder$^3$}   
\affil{$^1$SAAO, P.O.Box 9,
                Observatory, 7935, Cape Town, South Africa \\
$^2$Tuorla Observatory, 
University of Turku, FI-21500 Piikki\"o, Finland\\
$^3$AAO, P.O.Box 296, Epping, NSW 1710, Australia}  

\begin{abstract} 

Near infrared adaptive optics observations are crucial to be able to 
interpret kinematic and dynamical data and study star formation properties 
within the often extremely dusty interacting luminous IR galaxies 
(LIRGs). NIR AO data are also needed to find supernovae in their bright 
and dusty central regions and to fully characterize the young stellar 
clusters found in these kinds of systems. 
We have used AO in the $K$-band to survey a sample of LIRGs at $0.1$ 
arcsec (30 to 100 pc) resolution. The data are merged with SALT and AAT 
spectroscopic follow-up and HST and Spitzer archival imaging.
The first AO detected SNe are reported as well as details of the first studied
LIRGs. One LIRG showed an unexpected third component in the 
interaction, which moreover turned out to host the most active star formation. 
Another target showed evidence in the NIR of a very rare case of leading spiral
arms, rotating in the same direction as the arms open. 

\end{abstract}


\section{Introduction}   

In the local Universe luminous and ultra-luminous IR galaxies are a minority
(LIRGs have $L_{IR} = 10^{11} - 10^{12} L_{\odot}$ and ULIRGs
$L_{IR} > 10^{12} L_{\odot}$; \citealp[see][for a review]{sanders}). 
They are very dusty objects 
with a mixture of on-going AGN activity and strong star formation (SF), 
and hence
must also be locations of frequent core-collapse supernovae (CCSNe).  They 
are mostly interacting and merging gas-rich spirals. 
At higher redshifts they start to dominate the universal SF, LIRGs and
ULIRGs at $z\sim1$ and $z\sim2$, respectively, \citep[][]{lefloch}.  
It was proposed early on that they may be a link in an evolutionary sequence
between spiral galaxies and ellipticals, via mergers, obscured AGN and QSOs.
With this connection to early galaxy formation and evolution and to key galaxy
transformational processes, the local (U)LIRG population can be thought of as 
a laboratory to study in detail processes such as SF triggering and starburst 
vs.\ AGN interplay and black hole growth.

The picture has become more complicated however.
It is not clear that higher redshift (U)LIRGs have the same 
characteristics, apart from their high bolometric luminosity,
as the ones at lower redshifts.  Recent studies have shown that 
a large fraction of 
LIRGs at $z\sim0.6$ are quite regular disks without evidence 
for strong interactions \citep[e.g.][]{melbourne} and that higher 
redshift ULIRGs have much colder dust temperatures than local examples
\citep[e.g.][]{symeonidis}, temperatures that are closer to local cirrus 
dominated quiescent spirals. 

Most of the studies and surveys of IR-luminous galaxies over the past two
decades have concentrated on the most luminous and striking population, the 
ULIRGs.  
With the realization that the higher redshift ULIRGs 
actually have many physical characteristics (dust properties, SEDs, 
morphologies) more similar to local less-luminous IR galaxies closer to 
$L_{IR} \sim 10^{11} L_{\odot}$, larger detailed surveys of this latter 
class are well motivated.

\section{Observations}

What has been missing from local LIRG, as opposed to ULIRG,
surveys are very high spatial resolution imaging data in the NIR
to match optical HST data already in the archives. 
We have undertaken a $K$-band adaptive optics survey of local LIRGs
with $\sim0.1$\arcsec\ resolution.
Our pilot study was a 2-epoch survey with VLT/NACO and natural guide
stars of a sample of a dozen
southern LIRGs at $<200$ Mpc. 
We now have an on-going multi-epoch survey with Gemini ALTAIR/NIRI 
using laser guide stars.  This sample consists of 8 LIRGs at
$<120$ Mpc. The combined multi-epoch $K$-band images reach down to $K\sim22$
mag.  The observations are aimed at both identifying
CCSNe in the highly obscured central regions of LIRGs and to study the
LIRGs themselves in conjunction with spectroscopic follow-up observations
and archival optical HST and mid-IR Spitzer data. 
While the main statistical results are awaiting the completion of the Gemini
survey and more follow-up spectroscopy, we have already detected new CCSNe 
in both surveys, and followed up two of the LIRGs 
from the VLT survey in detail. These will be discussed below.

\section{The Bird: a triple merger}

\begin{figure}
  \plotone{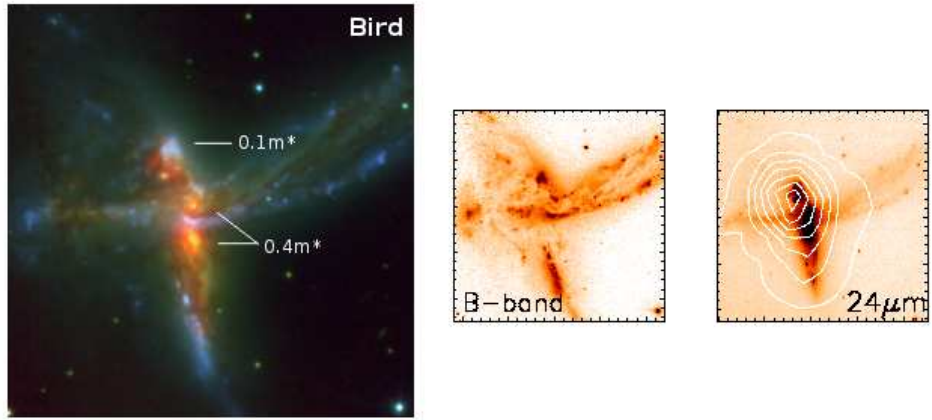}
  \caption{Left: IRAS~19115-2124 as a $BIK$
3-colour image. Masses of the nuclei in terms of $m_{\star}$ are indicated. 
Middle: 
the HST/ACS $B$-band highlights the large amount of dust 
which completely blocks from view the brightest NIR nucleus of the system.
Right: a contour overlay of Spitzer 24 $\mu$m emission on our NACO $K$-band 
image: the Head component dominates the SF.
}
  \label{figbird}
\end{figure}

NACO imaging of IRAS~19115-2124 was combined with Southern African
Large Telescope (SALT; \citealp{saltdod}) observations and HST/ACS and Spitzer
archival data for a detailed case study \citep{bird} -- 
we dubbed the system the Bird galaxy because of the obvious
resemblence (Fig.~\ref{figbird}).
The NIR imaging pinpointed the mass distribution in the system, which we 
then proceeded to study with long-slit spectroscopy ($R\approx2800$) 
in the range $\lambda$ = 5800 to 7200 \AA\ at various slit position angles. 
We determined line ratios, kinematics, velocity curves and 
dispersions, and mass estimates based on both spectral data and
$K$-band M/L estimates.
Three independent kinematical components were found
rather than the typical two interacting spirals in LIRGs. 
The ``Heart'' and ``Body'' components are the more massive ones, both 
$\sim 5 \times 10^{10} M_{\odot}$; the first is a spiral disk 
galaxy and the latter has a surface brightness 
profile intermediate between spirals and ellipticals. 
The ``Head'' is an irregular with $<1/4$ of the mass of the main 
components. It should be
stressed that the Head component can only be identified from the
NIR imaging and the Body is nearly completely obscured in the optical.  
The tidal tails, the Wings of the Bird, extend to 20 kpc.

A strong NaD doublet absorption feature in the SALT spectrum shows 
blue-shifted components interpreted as cool gas outflowing from 
the system, while blueshifted emission line ratios show signs of shock-heating.

A SFR of 190 ${\rm M}_{\odot} \ {\rm yr}^{-1}$ 
was derived from the optical-FIR SED.
Interestingly, the smallest component, the Head, 
is the one forming stars most rapidly, accounting for approximately 2/3 of the
24 $\mu$m flux of the whole system (Fig.~\ref{figbird}). 
Given its high offset velocity of $\sim400$
km/s with respect to the systemic velocity, we believe the Head must be on its
first encounter with the other two galaxy nuclei.  The more massive nuclei have
bar structures evident, which are thought to efficiently funnel gas to feed
central starbursts.  Interesting questions remaining to be answered include 
why the dominant SF is not at the central location,
and when and why was the current burst of SF in the Head triggered.

\section{Leading arms?}

\begin{figure}
  \plotone{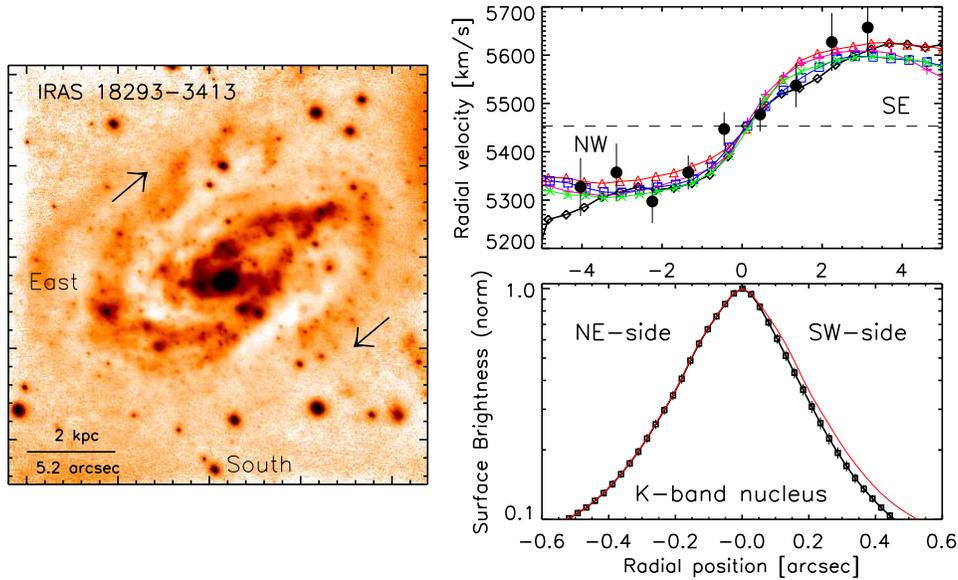}
  \caption{Left: IRAS~18293-3413 NACO $K$-band image. AAT NIR spectroscopy was 
taken along the major axis (SE to NW) and the rotation curve is shown at top 
right as curves for line emitting gas and as circles for CO-band absorption.
Bottom right: The black curve is a surface profile over the $K$-band bulge 
along the {\em minor axis}  The red thinner line is the the NE-side profile 
mirrored on the SW side, showing that the latter is more affected by extinction,
suggesting it is the nearer side to the observer resulting in the leading arm 
configuration. See \citet{leading} for discussion.}
  \label{figlead}
\end{figure}

IRAS~18293-3413 has been followed up with NIR spectroscopy
at the AAT and with archival data \citep{leading}.  This system also was
found to have a minor component, this time of less than 1/15 relative mass
and with a high relative velocity of $\sim500$ km/s.  The striking 
surprise here were the spiral arms of the main component, which are 
visible only in the $K$-band images.
Judging the orientation of the galaxy in space from extinction arguments
(Fig.~\ref{figlead}, right) and combining this with the direction of rotation 
from spectra, we arrive at the conclusion that the spiral arms move in the 
same direction that they open up, i.e.\ IRAS~18293-3413 is a leading arm spiral.

\begin{figure}
  \plotone{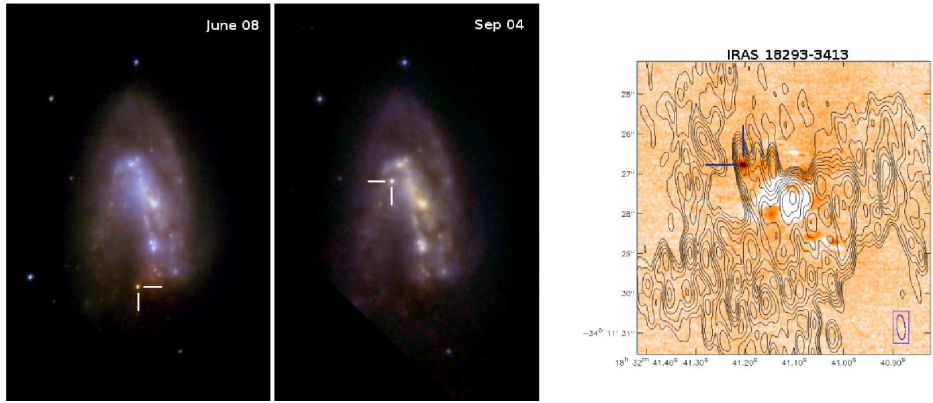} 
  \caption{IRAS~17138-1017 observed with Gemini AO in the NIR
bands at left, and an archival HST/NICMOS image in the middle. SN2008cs is 
seen in the lower part of the galaxy in the Gemini image 
and a ``historical'' SN2004iq 
left of the nuclear regions in the NICMOS image. Right:
VLA contour overlay on our subtracted $K$-band NACO data of IRAS~18293-3413.
SN2004ip is visible both in the NIR and radio wavelenghts.}
  \label{figsn}
\end{figure}

If the result holds up with more detailed data 
this would be one of only 2 or 3 convincing candidates of leading arm 
spirals \citep[][and references therein]{byrd,grouchy}. 
This kind of galaxy is not forbidden
by theory, and in fact simulations have shown that some retrograde encounters
should produce them \citep{thomasson} and that they would have implications
for e.g.\ the dark matter halo mass of spirals. 

\section{Searching for obscured core-collapse Supernovae}

The large SFRs in LIRGs are expected to result in CCSNe 
at rates a couple of orders of magnitude higher than
in ordinary field galaxies. 
This SN population is missed by optical ground based surveys because most of 
them are expected to occur in the bright nuclear regions, and are also 
likely to be heavily dust-obscured. NIR AO observations greatly increase 
their detectability with regard to 
both extinction and spatial resolution. The detected CCSNe can be used
as a direct and independent way to probe the SFR in galaxies 
both locally and in the distant Universe 
\citep[e.g.][]{dahlen04}.
  
Our first SN discovery SN2004ip \citep{seppo07} 
from the VLT/NACO sample was also the first ever AO-assisted SN discovered. 
Subsequent radio observations confirmed its CCSN nature 
(\citealp{miguel07}; Fig.~\ref{figsn} right). 
SN2004ip at a projected distance of 1.4'', or 500 pc, from the nucleus of 
IRAS 18293-3413, is among the closest SNe detected (in IR) to a LIRG nucleus.

Observations with Gemini ALTAIR/NIRI of IRAS 17138-1017 have so
far produced two SN discoveries (Fig.~\ref{figsn}, left \& centre; 
\citealp{kankare08}). 
SN2008cs is located at 4.2", or 1.5kpc, projected distance from the 
nucleus.  We obtained follow-up observations in $JHK$ bands which
are consistent with a core-collapse event suffering from a very high host 
galaxy extinction of $A_V \approx 16$ mag, the highest yet measured for any SN.
The CCSN nature of SN 2008cs was also confirmed by radio observations.
The ``historical'' SN 2004iq \citep{kankare08b} detected in the HST 
images is located 660 pc from the nucleus and suffers from a lower extinction.

\section{Super Stellar Cluster candidates}

All the LIRGs in our sample have many obvious point sources embedded in the
galaxies, likely young (super) stellar clusters (SSCs), which are expected 
to frequent strongly star forming galaxies. 
As seen in Fig.~\ref{figlead}, the spiral arms of 
IRAS~18293-3413 are ubiquitiously traced by 
SSC candidates in the NIR. Systems with stronger 
SFRs should have more SSCs, and it is more probable to 
find brighter and more massive SSCs in stronger starbursts \citep{larsen}. 
Indeed, the brightest SSC candidates in the Bird 
and IRAS~18293-3413 galaxies 
follow very well the SFR vs. brightest-$M_V$-cluster 
relation of \citet{bastian}. 

\begin{figure}
  \plotone{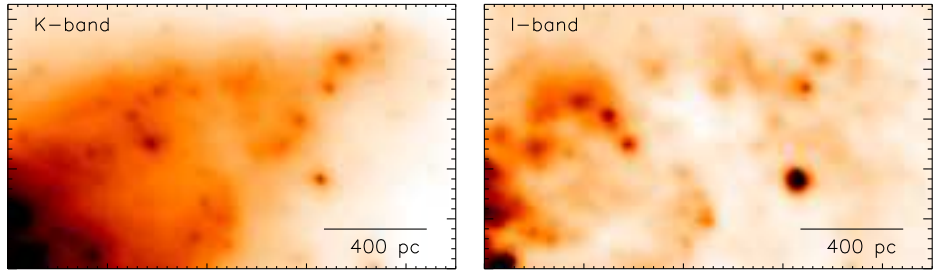}
  \caption{A region of 4.5''x2'' just North-West of the 
nucleus of IRAS~18293-3413 showing many SSC candidates. 
Our VLT/NACO data is on the left and HST/ACS $I$-data on the right.}
  \label{figssc}
\end{figure}

SSC characteristics and their luminosity functions have not been studied 
very much in the NIR yet. 
Our NIR AO-observations provide an excellent dataset to characterize 
the NIR SSC population for comparison with the optical studies, to see whether
LFs and dependencies of global galaxy parameters are similar.
As shown in Fig.~\ref{figssc} there are many
SSCs seen in $K$-band not seen at ACS $I$-band, and also SSCs at $I$-band which
are weak or not detected in the NIR, 
suggesting both heavy dust obscuration and age differences.
A full comparison of the $K$-band SSCs in our dozen or so LIRGs with optical 
SSC data and also with MIR detected deeply embedded clusters 
should be intriguing.

\section{Summary}   

It is absolutely critical to perform AO-assisted NIR observations to detect
what is really going on within the very dusty and often chaotic
(U)LIRGs. Otherwise whole progenitor nuclei in the interacting 
systems may be missed, and histories of the interactions and star formation
might be misinterpreted.  The data are also crucial in detecting a 
population of CCSNe in high SFR galaxies.

Two LIRG case studies were presented from an ongoing survey. SNe were detected,
as expected, and the new parameter space of resolution and wavelength provided
surprises in the form of ``extra'' nuclei and unusual spiral patterns. 
Follow-up studies of other LIRGs and their SSCs are continuing.

\acknowledgements 
PV thanks the conference organizers for a very interesting meeting and 
for the opportunity to present our survey. The rest of the ``SNe in LIRGs'' team
is warmly thanked for great work, especially Erkki Kankare and Miguel 
P\'erez-Torres regarding material presented here, and Travis Rector for the 
colour figures in Fig.~\ref{figsn}.


\end{document}